\documentclass[prb,preprint]{revtex4-1} 
\usepackage{amsmath}  
\usepackage{amsfonts} 
\usepackage{graphicx} 
\usepackage{subfigure}
\usepackage[dvipsnames]{xcolor}
\usepackage{ulem} 
\usepackage{hyperref}

\begin{document}

\title{The connection between a classical vibrating drumhead\\ and the masses of glueballs}


\author{Thales Azevedo}
\email{thales@if.ufrj.br}
\author{Henrique Boschi-Filho}
\email{boschi@if.ufrj.br}
\affiliation{Instituto de F\'isica, Universidade Federal do Rio de Janeiro, \\ Av. Athos da Silveira Ramos 149, 21941-909, Rio de Janeiro, Brazil}


\date{\today}

\begin{abstract}
The powerful techniques of holographic quantum chromodynamics (QCD) can be employed in the investigation of glueballs---composite particles made solely of gluons, the strong nuclear force mediators. In particular, the so-called hardwall model yields    predictions for the values of the masses of various glueball states, which are related to the solutions of the differential equations of the model.
 It turns out that those equations are essentially the same as the ones governing the vibrations of a circular membrane like that of a drumhead, which may serve as extra motivation for studying the dynamics of such an object as an undergraduate physics student.

\end{abstract}

\maketitle 

\section{Introduction}

 Whether string theory provides an accurate description of our universe or not, it has been the birthplace of several interesting ideas with several applications. For instance, the work of Kawai, Lewellen, and Tye on the relation between open and closed string amplitudes\cite{KLT} was the starting point of what became known as the ``double-copy'' framework, in which gravity observables (including those relevant to gravitational-wave astronomy) are computed from two copies of gauge theory scattering amplitudes---see  e.g. Refs. \cite{Bern:2019prr, Bern:2022wqg, Adamo:2022dcm, Mafra:2022wml, Lee:2025akf} for reviews. 

 Another example of a highly successful ``spin-off'' of string theory is the AdS/CFT correspondence,\cite{Maldacena:1997re, Witten:1998qj, Gubser:1998bc, Aharony:1999ti} a duality between a conformal field theory (CFT) in a flat, four-dimensional spacetime (without gravity) and a string theory in a five-dimensional Anti de Sitter (AdS) spacetime.\footnote{Technically, an AdS${}_5 \times \mathrm{S}^5$ spacetime, but the five-sphere $\mathrm{S}^5$ will play no role in our discussion.} One of the reasons for the success of the AdS/CFT correspondence is that it relates the strong-coupling regime on one side to the weak-coupling regime on the other side, effectively allowing perturbative computations of what would naively be regarded as non-perturbative quantities. 
 The AdS/CFT correspondence implies that  information from a higher-dimensional theory can be encoded in a lower-dimensional one, much like information from a three-dimensional object can be encoded in a two-dimensional hologram. For this reason, it is known as a holographic duality. 

 An active research field consists of using AdS/CFT-inspired techniques to study quantum chromodynamics (QCD), the gauge theory describing interactions of quarks and gluons. Since QCD is not a CFT, it is not possible to directly apply   AdS/CFT in order to compute QCD observables. Nevertheless, there are ways of adapting the correspondence to handle---at least approximately---gauge theories which are not conformally invariant. In the specific case of QCD, those attempts are known as holographic QCD or AdS/QCD models.

 In one of the simplest AdS/QCD models, conformal invariance is broken by artificially introducing an impassable barrier somewhere in the AdS space. In this so-called ``hardwall'' model,
 \cite{Polchinski:2001tt, Boschi-Filho:2002xih, Boschi-Filho:2002wdj, deTeramond:2005su, Erlich:2005qh, Boschi-Filho:2005xct} the masses of different QCD states are related to the position of that barrier in AdS. More precisely, the fields in AdS---corresponding to states in QCD---have to satisfy certain boundary conditions which, in turn, determine the values of the masses of QCD states.
 
 In particular, as we will explore in this paper, a scalar field in AdS corresponds to a scalar glueball state in QCD, i.e. a composite particle made solely of gluons, the strong nuclear force mediators. Glueballs have been  experimentally elusive, but recently one candidate has been detected\cite{BESIII:2024ein} with statistical significance greater than 5$\sigma$, to be confirmed by other research groups.

 When we impose a homogeneous Dirichlet boundary condition on the scalar field in AdS space, the  problem of finding glueball masses very much resembles the determination of the normal modes of a circular vibrating membrane, like that of a drumhead. Therefore, the ability to solve this typical undergraduate physics exercise comes in handy when investigating those nontrivial properties of QCD,  which gives students an extra reason to understand the concepts and techniques involved.

The remainder of this paper is organized as follows. In section~II, we review the solution of the classical circular membrane problem. In section~III, we briefly introduce the AdS/CFT correspondence and describe succinctly the equations obeyed by a scalar field in AdS space. In section IV, we introduce the main ingredients of the hardwall model and explain how it yields predictions for the masses of glueballs. In section V, we present our conclusions.

\section{Normal modes of a circular membrane}\label{membrane}

Here we briefly review, for pedagogical reasons, the problem of free vibrations of a circular membrane. A detailed discussion of this topic can be found in texts such as \cite{Jackson1962, Courant&Hilbert, Butkov1968, Boas1983, Wyld1999, Riley2006, Hayek2010, Borden2020}. According to Ref.~\cite{Hassani1999}, Leonhard Euler was the first to solve this problem. 

The problem is as follows. Consider a circular tensile membrane of radius $a$, similar to that of a drumhead. Under the usual assumptions of small oscillations (allowing linearization) and no external forces, one finds---by applying Newton's second law to an element of the membrane---that the function $u$ describing   displacements from the equilibrium satisfies the two-dimensional wave equation, i.e.
\begin{equation}\label{eq:2dwave}
   \frac{1}{v^2} \frac{\partial^2}{\partial t^2} u(\rho,\varphi,t)  = \nabla^2 u(\rho,\varphi,t) \,, \quad 0 < \rho < a\,,\quad -\pi < \varphi < \pi\,,\quad t>0\,, 
\end{equation}
where $\rho,\varphi$ are the usual polar coordinates and $v$ is the phase velocity, related to the surface tension of the membrane. Since we are modeling a drumhead, we also impose that the points on the boundary of the membrane never move, 
\begin{equation}\label{Dirichletmembrane}
    u(a,\varphi,t) = 0\,,\quad -\pi < \varphi < \pi\,,\quad t>0\,,
\end{equation}
besides the sensible conditions of $2\pi$-periodicity in $\varphi$ and analyticity as $\rho\to0$.

Now,  to find the normal modes of this vibrating membrane, it suffices to look for solutions of eq.~(\ref{eq:2dwave}) of the form
\begin{equation}
    u_\omega(\rho,\varphi,t) = f_\omega(\rho,\varphi) e^{i\omega t}\,, \quad \omega \in \mathbb{R}\,.
\end{equation}
Substituting this expression into the wave equation, we get
\begin{equation}\label{2dHelmholtz}
    \nabla^2 f_\omega(\rho,\varphi) + k_\omega^2 f_\omega(\rho,\varphi)  = 0\,,
\end{equation}
 where we defined $k_\omega\equiv |\omega|/v$. Now we can write $f_\omega(\rho,\varphi)=P_{\omega,N}(\rho)e^{\pm iN\varphi}$, with $N \in \mathbb{Z}_+$. Note that this implies $f_\omega(\rho,\varphi)$ is $2\pi$-periodic. Then, using the expression for the Laplacian in polar coordinates, eq.~(\ref{2dHelmholtz}) yields
\begin{equation}\label{Besselmembrane}
    \rho^2 \frac{\mathrm{d}^2}{\mathrm{d}\rho^2}P_{\omega,N} + \rho \frac{\mathrm{d}}{\mathrm{d}\rho}P_{\omega,N} + \left(k_\omega^2\rho^2 - N^2\right)P_{\omega,N} = 0\,.
\end{equation}

Equation~(\ref{Besselmembrane}) can readily be recognized as a particular instance of the (generalized) Bessel differential equation \cite{Boas1983, Hayek2010} 
\begin{equation}\label{Besselgeneral}
    x^2 y''(x) + (1-2\alpha)xy'(x) + \left[ \left(\beta\gamma x^{\gamma} \right)^2 + \alpha^2 -\nu^2\gamma^2\right]y(x) = 0\,,
\end{equation}
 whose general solution is given by
\begin{equation}\label{Besselsolution}
    y(x) = c_1\, x^\alpha J_\nu(\beta x^\gamma) + c_2\, x^\alpha Y_\nu(\beta x^\gamma)\,,
\end{equation}
where $c_1,c_2$ are constants and $J_\nu$ and $Y_\nu$ are, respectively, the Bessel and Neumann functions of order $\nu$. Comparing eqs.~(\ref{Besselmembrane}) and (\ref{Besselgeneral}), it is easy to see that $\alpha=0$, $\beta = k_\omega$, $\gamma = 1$ and $\nu=N$. Hence, from eq.~(\ref{Besselsolution}) we obtain
\begin{equation}
    P_{\omega,N}(\rho) = c_1\,J_N(k_\omega\rho) + c_2\,Y_N(k_\omega\rho)\,.
\end{equation}
Now, since the Neumann function is not analytic at $\rho=0$, we must set $c_2=0$. Moreover, we must impose the homogeneous Dirichlet boundary condition at $\rho=a$ (eq.~(\ref{Dirichletmembrane})), which leads to
\begin{equation}
     P_{\omega,N}(a) = 0 \;\Longrightarrow \; J_N(k_\omega a) = 0 \; \Longrightarrow \; k_\omega  = \frac{\xi_{N,\ell}}{a}\quad (\ell \in \mathbb{N})\,,
\end{equation}
where $\xi_{N,\ell}$ denotes the $\ell$-th zero of $J_N$.

Finally, we find that the normal modes of the circular membrane are given by
\begin{equation}
    u^\pm_{N,\ell}(\rho,\varphi,t) = A_{N,\ell}\, J_N(\xi_{N,\ell}\rho/a) e^{\pm iN\varphi} e^{i\omega_{N,\ell}t}\,,
\end{equation}
where $A_{N,\ell}$ is some convenient normalization constant and the allowed resonant frequencies are $\omega_{N,\ell}=\pm v\xi_{N,\ell}/a$,  since $k_\omega=|\omega|/v$. In general, the vibration of the membrane is described by a superposition of its normal modes, whose precise form depends on the initial conditions.

\section{Brief review of A$\mathbf{d}$S/CFT and Scalar fields in A$\mathbf{d}$S Space}
\label{review}

Briefly, the AdS/CFT correspondence\cite{Maldacena:1997re} is the statement that a superstring theory defined in a ten-dimensional spacetime AdS${}_5\times \mathrm{S}^5$ and a supersymmetric Yang-Mills (SYM) field theory in ordinary four-dimensional Minkowski space are equivalent to each other. The five-dimensional Anti-de Sitter space (AdS${}_5$)  is a space of constant negative curvature, while the five-sphere S${}^5$ is a space of constant positive curvature. This highly nontrivial duality can partly be motivated through symmetry arguments, since the isometry group of AdS${}_5$ coincides with the conformal symmetry group of that four-dimensional SYM theory. Since this duality relates theories in spacetimes with different dimensions, it is often referred to as holography, in   allusion to the optical analog of encoding three-dimensional objects on two-dimensional surfaces. 

According to AdS/CFT, the strong-coupling regime of SYM is dual to weakly coupled superstrings, which can roughly be treated as a field theory in AdS${}_5\times \mathrm{S}^5$ (to leading order in the string-length parameter). Since AdS${}_5\times \mathrm{S}^5$ is a curved spacetime, we employ the usual language of Einstein's general relativity theory.

For our purposes here, it will suffice to ignore the five-sphere and focus on  AdS${}_5$. The metric of this space, $g_{MN}$ ($M,N = 0$ to 4), can be written as
    \begin{equation}\label{metric}
        \mathrm{d}s^2= g_{MN}\mathrm{d}x^M \mathrm{d}x^N=\frac{R^2}{z^2}(\eta_{\sigma\tau}\mathrm{d}x^\sigma \mathrm{d}x^\tau+\mathrm{d}z^2)\,,
    \end{equation}
    where $R$ is the AdS${}_5$ radius,  $\eta_{\sigma\tau}$ ($\sigma,\tau=0$ to $3$) is the Minkowski space metric in Cartesian coordinates $x^\sigma$ (with $-,+,+,+$ signature) and $z\equiv x^4$ is a radial coordinate, usually called the fifth coordinate of the AdS${}_5$ space, defined in the interval $[0,\infty)$. In these coordinates, it is possible to show that the boundary of AdS${}_5$ corresponds to   $z=0$. 

The glueball states in which we are interested live in the four-dimensional boundary of AdS${}_5$. The AdS/CFT correspondence tells us that  scalar glueball states are holographically described by scalar fields in AdS${}_5$ space. 
The dynamics of a scalar field $\phi(z,x^\sigma)$ with mass $m_5$ in the AdS${}_5$ space defined by the metric (\ref{metric}) is governed by the Klein-Gordon equation,\footnote{We adopt a system of natural units such that $c=\hbar=1$.} $g^{MN}\nabla_M\nabla_N\phi - m_5^2\phi= 0$, which yields
    \begin{equation}\label{kgequation1}
        z^{5}\partial_z(z^{-3}\partial_z\phi)+z^2\eta^{\sigma\tau}\partial_\sigma\partial_\tau\phi-(m_5R)^ 2\phi=0\,.
    \end{equation}
    Now, for each fixed value of $z$, let $\widehat\phi(z,k^\sigma)$ be the Fourier transform of   $\phi(z,x^\sigma)$. Thus,
    \begin{equation}
        \phi(z,x^\sigma)=\int\frac{\mathrm{d}^4k}{(2\pi)^4}e^{ik\cdot x}\widehat\phi(z,k^\sigma)\,.
    \end{equation}
Substituting the above form of $\phi(z,x^\sigma)$ into equation~(\ref{kgequation1}) and using the orthogonality of complex exponentials, we obtain
    \begin{equation}\label{eq.phihatedo}
        z^{2}\partial^2_z\widehat\phi -3z \partial_z \widehat\phi -k^2z^2\widehat\phi -(m_5R)^2\widehat\phi=0\,.
    \end{equation}

Comparing Eq.~(\ref{eq.phihatedo}) with the general form Eq.~(\ref{Besselgeneral}), we find in this case that they match if $\alpha=2$, $\beta=\sqrt{-k^2}$, $\gamma=1$ and $\nu = \sqrt{4+(m_5R)^2}$. Concerning the terms that involve square roots, two comments are in order. First, note that from a four-dimensional point of view, we should indeed have $k^2\equiv k_\sigma k^\sigma<0$ for a massive particle (given our choice of metric signature), so $\beta$ is a positive real number. Second, if we impose that $\nu\in\mathbb{R}$, then the mass of the scalar field in AdS${}_5$ must satisfy
\begin{equation}
        m_5^2\geq-\bigg(\frac{2}{R}\bigg)^2,
    \end{equation}
which is known as the Breitenlohner-Freedman bound.\cite{Breitenlohner:1982jf}

Now, given the values of $\alpha,\beta,\gamma,\nu$ that we found, the solution to Eq.~(\ref{eq.phihatedo}) is (see Eq.~(\ref{Besselsolution}))
\begin{equation}
    \label{J+Y}\widehat\phi(z,k^\sigma) = \widehat A(k^\sigma)\,z^2J_\nu(\sqrt{-k^2} z) + \widehat B(k^\sigma)\,z^2Y_\nu(\sqrt{-k^2} z)\,,
\end{equation}
in terms of Bessel and Neumann functions of the fifth coordinate $z$ of the AdS${}_5$ space. Since we want to describe glueballs in terms of these solutions, we are going to discard the Neumann term, which is related to non-normalizable modes.


\section{Glueball masses in the Hardwall model}


The AdS/CFT duality cannot directly be applied to the study of non-perturbative QCD properties because QCD is not a conformal field theory (CFT). Indeed, CFTs do not have any mass scale, for example, which clearly is not the case for QCD, since protons and neutrons acquire their masses mostly from the strong interaction. 

To circumvent that obstacle, the hardwall model\cite{Polchinski:2001tt, Boschi-Filho:2002xih, Boschi-Filho:2002wdj} 
introduces a cutoff in the AdS${}_5$ space at some value of the fifth coordinate $z=z_{\rm max}$, so that now the range of this coordinate is reduced to the interval $[0,z_{\rm max}]$. Accordingly, one needs to impose a boundary condition at $z=z_{\rm max}$, which we will consider to be of the Dirichlet kind. The introduction of $z_\mathrm{max}$ implies that now there is a length scale in the model that breaks the conformal invariance of the dual field theory, allowing the description of massive states, for example.

Now, let us consider again the  solution in equation~\eqref{J+Y}. 
In the hardwall model, scalar glueballs are related to scalar fields in AdS${}_5$, and their masses   correspond to the possible values of $\sqrt{-k^2}$. Discarding the Neumann function   and imposing a homogeneous Dirichlet boundary condition at $z=z_\mathrm{max}$, we get
\begin{equation}\label{eq.Jglueball}
     \left.\widehat\phi(z,k^\sigma)\right|_{z=z_\mathrm{max}} = 0 \; \Longleftrightarrow \;   J_\nu(\sqrt{-k^2} z_\mathrm{max})=0\,.
\end{equation}
Note that this equation has exactly the same form as the one we solved to find the resonant frequencies of the circular membrane.

Indeed, there is a clear analogy between the resonant frequencies of the membrane and the masses of glueballs. Just as the membrane has a fundamental vibrational mode, related to the lowest zero of the Bessel function, the glueball has a ``ground state mass." From eq.~\eqref{eq.Jglueball}, we see that different masses $M^{(g)}$ (possible values of $\sqrt{-k^2}$)  satisfy
\begin{equation}
    M^{(g)}_{\nu,1}\,z_\mathrm{max}=\xi_{\nu,1}\,, \quad M^{(g)}_{\nu,2}\,z_\mathrm{max}=\xi_{\nu,2}\,, \ldots\,,
\end{equation}
i.e. the following ratios are constant,
\begin{equation}  \label{eq.massratios}  
    \frac{M^{(g)}_{\nu,1}}{\xi_{\nu,1}} = \frac{M^{(g)}_{\nu,2}}{\xi_{\nu,2}} = \cdots \,.
\end{equation}

Now, from the discussion in the previous section, we know that the order of the Bessel function ($\nu$) is related to the mass of the corresponding scalar field in AdS${}_5$ through
\begin{equation}
    \nu = \sqrt{4+(m_5R)^2}\,.
\end{equation}
So we still need to explain the relation between $m_5$, $R$ and the masses of glueballs.
In fact, depending on the spin and the conformal dimension of the dual operator, glueball states of different spins are associated with Bessel functions of different orders. For instance, in models discussed in the literature the lightest scalar glueball (spin 0) may be associated with a Bessel function of order \(\nu=2\) (corresponding to a massless scalar in AdS${}_5$, i.e. $m_5=0$), while higher-spin states map to other values of $\nu$. Then, for scalar glueballs ($\nu=2$), the relation 
\eqref{eq.massratios}, reduces to 
\begin{equation}  \label{eq.massratiosnu=2}  
    \frac{M^{(g)}_{2,1}}{\xi_{2,1}} = \frac{M^{(g)}_{2,2}}{\xi_{2,2}} = \cdots \,.
\end{equation}

Scalar glueballs are still not confirmed experimentally, but we can  compare our predictions with independent lattice-QCD data presented in Table~I, from ref. \cite{Lucini:2004my} (see also 
ref. \cite{Rinaldi:2021dxh} for a nice compilation of glueball masses). The lightest scalar glueball state is named $0^{++}$, since it has 0 spin, positive parity, and positive charge conjugation. Its first radial excitation is named $0^{++*}$, the second $0^{++**}$, etc. In the hardwall model, the mass of the $0^{++}$ state, $M^{(g)}_{2,1}$,  can be taken as an input from the lattice, with a value of 1475 MeV. Then, using the above relation, we calculate the masses of the radial excitations of the scalar glueball states, as shown in Table~I.  For example, the mass of the glueball state $0^{++*}$, corresponding to $M^{(g)}_{2,2}$, can be obtained from eq.~\eqref{eq.massratiosnu=2} as
\begin{equation}
    M^{(g)}_{2,2} = \frac{\xi_{2,2}}{\xi_{2,1}}M^{(g)}_{2,1} = \frac{8.4172}{5.1357} \times 1475 \;{\rm MeV} \approx 2418 \;{\rm MeV}\,,
\end{equation}
in agreement with the result shown in Table~I.

\begin{table}
    \centering
    \begin{tabular}{|c||c|c|c|c|}
    \hline
       States  & $0^{++}$ & $0^{++*}$ & $0^{++**}$ & $0^{++***}$\\
       \hline 
        Masses & $M^{(g)}_{2,1}$ & $M^{(g)}_{2,2}$ & $M^{(g)}_{2,3}$ & $M^{(g)}_{2,4}$ \\
        \hline 
        Lattice\cite{Lucini:2004my, Rinaldi:2021dxh} & 1475 $\pm$ 72 & 2755 $\pm$ 124 & 
        3370 $\pm$ 180 & 
        3990 $\pm$ 277\\
        \hline
        This work & 1475 $\pm$ 72 &  2418 $\pm$ 118 &  3337  $\pm$ 163 &   4250 $\pm$ 207
        \\
        \hline 
        Zeros of $J_2$ & 5.1357 & 8.4172 & 11.6198 & 14.7960 \\
        \hline 
    \end{tabular}
    \caption{Scalar Glueball states and masses in MeV as predicted by lattice QCD and this work. The lightest glueball state ($0^{++}$) mass from the lattice is taken as an input here. The other glueball states ($0^{++*}$, $0^{++**}$, $0^{++***}$) correspond to radial excitations. For clarity, we also present the first four zeros of $J_2(x)$ used in relation  \eqref{eq.massratiosnu=2} to calculate the scalar glueball masses from the hardwall model.}
    \label{tab:placeholder}
\end{table}


\section{Conclusions}

In physics, one often explores analogies in order to better understand or describe a given system. In other words, the use of analogies can bridge the gap between complex, unfamiliar physical situations and concepts that are already well-understood. 


In this paper,  we have shown how a simple classical mechanics problem helps visualize the  spectrum  of     a quantum system. Namely, we have used the hardwall model of AdS/QCD to explore the analogy between the masses of glueball states and the resonant frequencies of a circular membrane like that of a drumhead.
Indeed,  in the same way that a drumhead has a ``fundamental frequency" and higher ``overtones" based on its physical properties, glueballs have a ``ground state mass" and ``excited state masses." In this analogy, the mass of a glueball corresponds to the frequency of a specific vibrational mode of the membrane.

Furthermore, the classical and quantum problems are related by their
discrete spectra. Just as a membrane can only vibrate at specific, discrete frequencies (modes), the masses of glueballs are quantized. One can then use the known patterns of membrane vibrations to predict the mass ratios of different glueball states, as we have seen.


A detailed explanation of how the hardwall model determines glueball masses involves advanced concepts from general relativity and quantum field theory that are beyond the scope of a typical undergraduate course. However, we believe that merely mentioning this striking connection---linking the measurable vibrations of a macroscopic object to the inherent masses of subatomic particles like glueballs---can inspire students and underscore the importance of mastering core physics equations.

\vskip 1 cm

\noindent
{\bf Acknowledgements}\\
HBF is partially supported by Conselho Nacional de Desenvolvimento Cient\'{\i}fico e Tecnol\'{o}gico (CNPq) under grant  310346/2023-1, and Fundação Carlos Chagas Filho de Amparo à Pesquisa do Estado do Rio de Janeiro (FAPERJ) under grant E-26/204.095/2024.

\vskip 1 cm

\noindent
{\bf Data availability statement}

\noindent No new data were created or analysed in this study.


\end{document}